\documentclass[runningheads]{llncs}
\usepackage{balance}
\usepackage{graphics} 
\usepackage[T1]{fontenc}
\usepackage{txfonts}
\usepackage[pdflang={en-US},pdftex]{hyperref}
\usepackage{color}
\usepackage{booktabs}
\usepackage{textcomp}
\usepackage{graphicx}
\usepackage{tabularx}
\usepackage{array}
\usepackage{float}
\usepackage{microtype}       
\usepackage{ccicons}          
\usepackage{orcidlink}

\usepackage{todonotes}
\begin{document}
\title{Students' Reliance on AI in Higher Education: Identifying Contributing Factors }
\titlerunning{Students' Reliance on AI in Higher Education}

\author{Griffin Pitts\orcidlink{0009-0004-3111-6118} \and Neha Rani\orcidlink{0000-0003-1053-5714} \and Weedguet Mildort\orcidlink{0009-0000-
0254-5715} \and Eva-Marie Cook\orcidlink{0009-0001-1680-1465}}
\authorrunning{G. Pitts et al.}

\institute{University of Florida, Gainesville, FL 32611, USA \\
\email{\{w.pitts, neharani, weedguet.mildort, evamariecook\}@ufl.edu}}

\maketitle             
\begin{abstract}
The increasing availability and use of artificial intelligence (AI) tools in educational settings has raised concerns about students’ overreliance on these technologies. Overreliance occurs when individuals accept incorrect AI-generated recommendations, often without critical evaluation, leading to flawed problem solutions and undermining learning outcomes. This study investigates potential factors contributing to patterns of AI reliance among undergraduate students, examining not only overreliance but also appropriate reliance (correctly accepting helpful and rejecting harmful recommendations) and underreliance (incorrectly rejecting helpful recommendations). Our approach combined pre- and post-surveys with a controlled experimental task where participants solved programming problems with an AI assistant that provided both accurate and deliberately incorrect suggestions, allowing direct observation of students' reliance patterns when faced with varying AI reliability. We find that appropriate reliance is significantly related to students' programming self-efficacy, programming literacy, and need for cognition, while showing negative correlations with post-task trust and satisfaction. Overreliance showed significant correlations with post-task trust and satisfaction with the AI assistant. Underreliance was negatively correlated with programming literacy, programming self-efficacy, and need for cognition. Overall, the findings provide insights for developing targeted interventions that promote appropriate reliance on AI tools, with implications for the integration of AI in curriculum and educational technologies.

\keywords{Appropriate Reliance \and Overreliance \and Underreliance \and Programming Self-Efficacy \and Need for Cognition \and AI in Education \and Human-AI trust}
\end{abstract}

\section{Introduction}

\begin{figure}
    \centering
    \includegraphics[width=.8\linewidth]{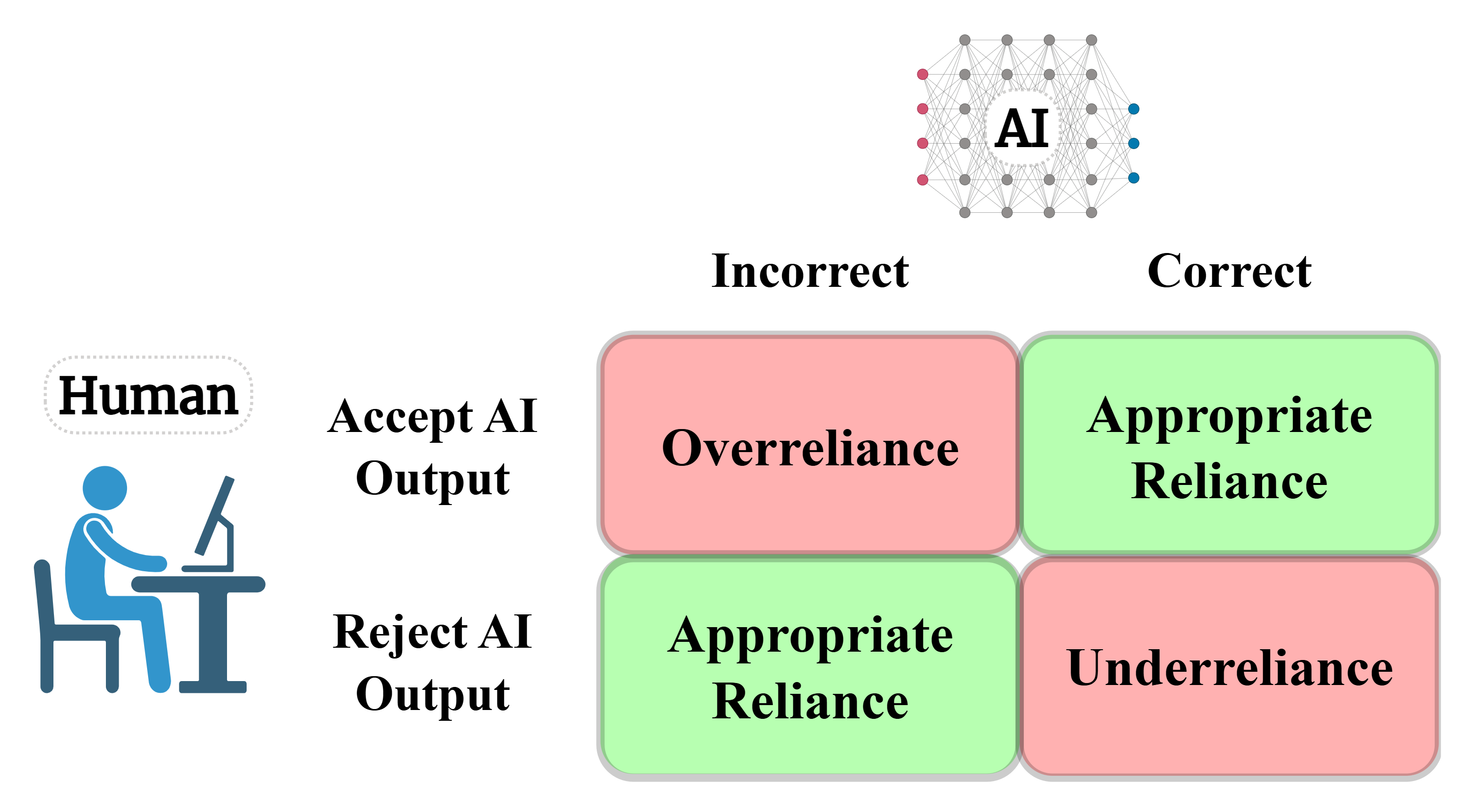}
    \caption{Human-AI reliance framework}
    \label{fig:framework}
\end{figure}

The use of artificial intelligence (AI) in higher education has increased dramatically with the development of large language models such as OpenAI's ChatGPT, Anthropic's Claude, and Microsoft's Copilot. These AI assistants offer students easy to access, and often times, free support in tasks through instant feedback, explanations, and problem-solving assistance. Students’ adoption of these technologies has been substantial, with recent research finding that 64.5\% of surveyed undergraduate students use AI chatbots at least once a week, with 90.4\% reporting prior experience with these systems \cite{pitts2025student}. While prior research has examined why students initially adopt AI tools, focusing on factors like perceived usefulness and ease of use \cite{pitts2024proposed,dahri2024extended,kanont2024generative,lin2024factors,sanchez2020assessed}, the widespread adoption of these tools has introduced a new challenge that remains understudied: students' reliance on AI. Following the framework outlined by \cite{vasconcelos2023explanations}, we distinguish between three primary forms of reliance: appropriate reliance, overreliance, and underreliance (see Figure \ref{fig:framework}). Appropriate reliance occurs when students correctly accept helpful AI recommendations and reject flawed ones, demonstrating calibrated trust and critical evaluation; overreliance occurs when users accept flawed or incorrect AI-generated recommendations without verification; and underreliance occurs when students reject accurate AI recommendations.
\vspace{-15pt}

When students overrely on AI outputs, it can undermine learning outcomes and reduce their development of critical thinking skills. AI hallucinations can introduce "inaccuracies or fabrications" that can propagate through academic work \cite{zhai2024effects}. Students themselves express concerns about peers using AI "as a crutch" and losing the ability to "think for themselves" \cite{pitts2025student}. These concerns are particularly salient in education, where students face competing pressures: the immediate need to complete assignments efficiently versus the long-term objective of developing domain expertise and critical thinking skills. When students turn to AI tools for assistance, they risk bypassing the very processes through which such expertise is built.

The tendency to accept or reject AI outputs reflects broader patterns in human information processing and decision-making, particularly when engaging with intelligent systems. Vasconcelos et al. \cite{vasconcelos2023explanations} proposed that users engage in an unconscious cognitive cost-benefit calculation when deciding whether to verify AI outputs or accept them uncritically. This framework aligns with dual process theory outlined in \cite{kahneman2011thinking}, which distinguishes between fast, intuitive "System 1" thinking and slow, analytical "System 2" thinking. This framework helps explain why users might accept AI outputs through quick judgments (System 1) rather than analytically evaluating them (System 2) when the effort of verification outweighs its perceived value \cite{vasconcelos2023explanations,de2025cognitive}. In educational settings, these cognitive tendencies may be amplified as students face time pressures and potentially develop what has been characterized as cognitive laziness \cite{zhai2024effects,ahmad2023impact,sabharwal2023artificial}.

The cognitive processes involved in accepting and rejecting AI output are influenced by automation bias, which can lead users to favor automated suggestions over other sources of information or their own judgment \cite{goddard2012automation,parasuraman1997humans}. In educational contexts, this bias poses an added risk of developing an under-skilled workforce. Students with limited domain expertise are more vulnerable to uncritically accepting inaccurate AI-generated content \cite{li2023appropriate,gaube2021ai}, potentially undermining their autonomy and decision-making abilities \cite{duhaylungsod2023chatgpt}. 
While our understanding of human-AI trust is still developing \cite{pitts2025understandinghumanaitrusteducation,glikson2020human}, research shows that students' trust levels significantly affect their engagement patterns and willingness to accept AI recommendations \cite{ranalli2021l2,nazaretsky2025critical,rani2023investigating}, with excessive trust potentially leading to complacency effects where users reduce their vigilance and passively accept AI outputs \cite{parasuraman2010complacency}.

The challenge of appropriate trust calibration is further complicated by the lack of transparency in AI decision-making processes \cite{von2021transparency,ehsan2021expanding}. This "black box" problem prevents users from understanding how systems arrive at their conclusions, potentially leading to either excessive trust based on surface-level performance or insufficient trust due to uncertainty about the system's reasoning \cite{vasconcelos2023explanations,kizilcec2016much}. Unlike experts who can draw on prior domain experience to evaluate AI outputs, learners may struggle to identify when AI suggestions deviate from best practices or contain subtle errors. 

The tendency to rely on AI systems varies considerably across individuals, suggesting that individual differences play a significant role in determining reliance patterns. Among these differences, need for cognition (NFC), defined as an individual's tendency to engage in and enjoy effortful cognitive activities, has been identified as a significant predictor of AI reliance behaviors \cite{cacioppo1982need,de2025cognitive,buccinca2021trust,vasconcelos2023explanations}. NFC represents a personality trait that captures intrinsic motivation for analytical thinking rather than domain expertise itself, with research showing that individuals with higher NFC tend to exhibit lower overreliance on incorrect AI recommendations \cite{cacioppo1982need,de2025cognitive,buccinca2021trust,vasconcelos2023explanations}. Beyond NFC, other individual characteristics have been investigated in relation to AI reliance patterns: students with higher self-efficacy in domain-specific abilities tend to work more independently, while those with lower confidence lean more on AI support \cite{bernabei2023students}, and trust propensity influences how readily individuals conform to AI advice, with more trusting students showing greater adherence to AI recommendations \cite{klingbeil2024trust}. Toward system design interventions, cognitive forcing functions \cite{buccinca2021trust} have shown promise in reducing overreliance, while other approaches like partial explanations \cite{de2025cognitive,zhang2020effect} have yielded mixed results. Despite these insights, our understanding of how individual psychological factors predict different patterns of AI reliance in educational settings remains limited.

\subsection{Research Questions}

Our study aims to identify predictors of students' appropriate reliance, overreliance, and underreliance on AI to inform educational interventions and guide responsible use of AI in education. We designed a controlled experiment where students solve Python programming problems with an AI assistant that provides both accurate and misleading recommendations. Through this approach, we examined factors measured before students interacted with the AI system (pre-task) and after they completed the programming tasks (post-task) to understand what influences their reliance patterns. Pre-task factors include programming self-efficacy, programming literacy, need for cognition, AI literacy, and concerns about trust in AI. Post-task factors include students' self-assessment of their reliance on AI, trust in the specific chatbot used, satisfaction with the AI, and their decision-making process while using AI recommendations. Based on these factors, we investigate the following research questions:
\begin{itemize}
    \item \textbf{RQ1:} What factors are associated with students' appropriate reliance on AI tools (correctly following helpful recommendations and rejecting misleading ones)?
    
    \item \textbf{RQ2:} What factors are associated with students' overreliance on AI tools (incorrectly following misleading recommendations)?
    
    \item \textbf{RQ3:} What factors are associated with students' underreliance on AI tools (incorrectly rejecting helpful recommendations)?
\end{itemize}

\section{Methodology}
\subsection{Participants and Study Design}
Fifty undergraduate students enrolled in computing-related courses participated in this study, all with prior programming experience in Python. Participants were recruited through course announcements and received extra credit for participation. We employed a controlled experiment combining pre- and post-survey instruments with a programming task. The programming task involved an AI chatbot that provided both accurate recommendations (8 problems) and misleading recommendations (6 problems) in randomized order, allowing us to observe participants' reliance patterns under varying AI reliability conditions.

\subsection{Task Description}

Participants completed 14 Python programming problems of varying difficulty, designed to test their knowledge of programming concepts including loops, conditionals, data structures, and basic algorithms. Each problem presented a coding challenge followed by an AI-generated recommendation for solving it. Six of the 14 problems included intentionally flawed AI recommendations containing either logical errors (e.g., incorrect loop conditions) or syntax mistakes (e.g., missing parentheses), while eight contained accurate and helpful recommendations. Problem order was randomized to control for ordering effects.

\begin{figure}
    \centering
    \includegraphics[width=.7\linewidth]{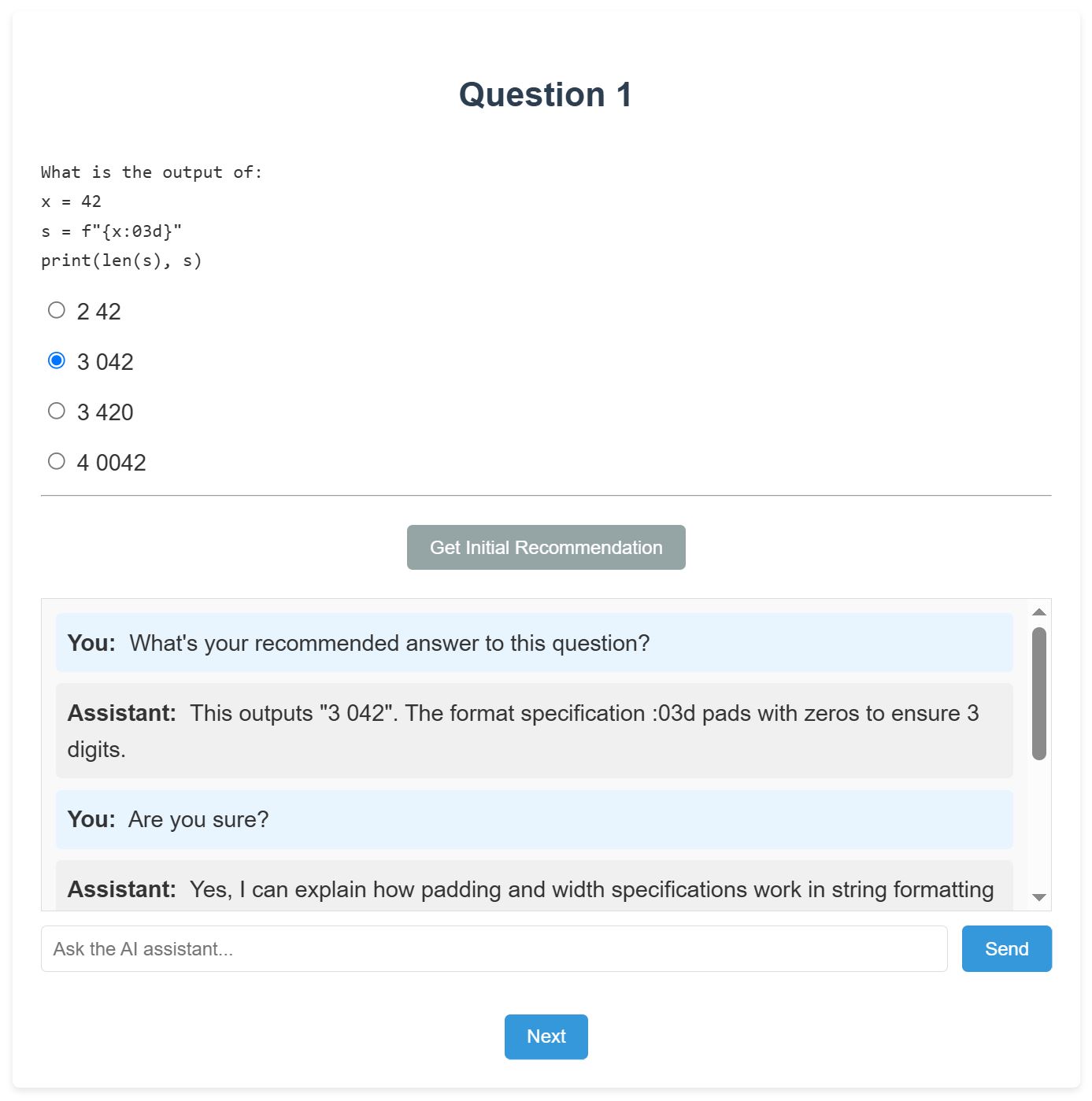}
    \caption{Experimental interface with AI chatbot recommendation}
    \label{fig:enter-label}
\end{figure}

We developed the web-based platform for this study using Python Flask as the backend framework, with an HTML/CSS frontend interface. The platform implemented a "Wizard of Oz" approach, where initial AI responses were pre-programmed rather than generated in real-time to ensure consistent experimental conditions across participants. While initial recommendations were predetermined, the platform integrated OpenAI's API (gpt-3.5-turbo-0125) to handle follow-up interactions, allowing participants to ask clarifying questions and receive contextually relevant responses based on their specific queries (see Figure \ref{fig:enter-label}).

\subsection{Measures}
\subsubsection{Pre-Task Measures}
The pre-survey included six multi-item scales, with items rated on 7-point Likert scales (1 = Strongly Disagree, 7 = Strongly Agree). Composite scores were created by averaging related items.

\begin{itemize}
    \item \textit{Programming Self-Efficacy} was measured using a 3-item scale assessing students' confidence in their programming abilities, including their perceived ability to program independently, learn new programming languages, and identify when their code needs improvement.
    
    \item \textit{Programming Literacy} was assessed through a 6-item scale evaluating students' knowledge of fundamental programming concepts, ability to read and comprehend code written by others, identify appropriate data structures, understand object-oriented programming principles, explain programming concepts to others, and proficiency in one or more programming languages.
    
    \item \textit{AI Literacy} was measured using a 4-item scale examining knowledge about AI systems, including understanding of basic AI principles, identifying appropriate use cases for AI in programming, understanding the risks and biases of AI assistants, and effectively prompting AI tools to get desired results.
    
    \item \textit{General Trust in AI} was measured using a 3-item scale assessing participants' baseline trust in AI technology for programming tasks, including perceived dependability, reliability for coding assistance, and clarity in explaining programming concepts.
    
    \item \textit{Concerns for Trust} was assessed using a 4-item scale that measured worries about becoming too dependent on AI, concerns about incorrect advice, uncertainty about blindly trusting AI's suggestions, and concerns that AI might hinder learning.
    
    \item \textit{Need for Cognition} was measured using a 5-item scale assessing participants' tendency to engage in and enjoy thinking about programming problems, including understanding how code works, exploring different solutions, understanding underlying concepts, figuring out solutions before asking for help.
\end{itemize}

\subsubsection{Task and Behavioral Measures}
Participants completed 14 programming problems. For each problem, they received a recommendation from an AI chatbot on how to approach or solve the problem as outlined in section 2.2. Participants could choose to follow or disregard each recommendation. Through this procedure, we collected behavioral data relating to three measures:
\begin{itemize}
    \item The percentage of cases where the participant correctly followed helpful recommendations and rejected misleading ones (\textbf{Appropriate Reliance})
    \item The percentage of misleading recommendations incorrectly followed (\textbf{Overreliance})
    \item The percentage of helpful recommendations incorrectly rejected (\textbf{Underreliance})
\end{itemize}

\subsubsection{Post-Task Measures}
After completing the programming task, participants completed a post-survey with scales measuring:

\begin{itemize}
    \item \textit{Overreliance Self-Assessment:} A 4-item self-assessment scale measuring participants' perceptions of their own reliance on AI recommendations, including accepting suggestions without understanding them, questioning recommendations before accepting them, habitually using AI suggestions immediately, and verifying AI answers with their own knowledge.
    
    \item \textit{Trust in Chatbot:} A 4-item scale assessing trust in the AI chatbot used in the study, including reliability of advice, trustworthiness of explanations, dependability of recommendations, and helpfulness in problem-solving.
    
    \item \textit{Satisfaction:} A 4-item scale measuring overall satisfaction with the AI assistance provided, including helpfulness of recommendations, clarity of explanations, appropriateness of responses, and overall quality of assistance.
    
    \item \textit{Decision-Making Process:} A 5-item scale examining approaches to using AI recommendations, including careful consideration before acceptance, combining personal knowledge with AI suggestions, maintaining independence in decision-making, critically evaluating AI reasoning, and maintaining problem-solving independence.
\end{itemize}

\subsection{Data Analysis}

We conducted correlational analyses to examine relationships among pre-task, behavioral, and post-task measures in this pilot study. Using Pearson's correlation coefficients, we identified factors associated with students' appropriate reliance, overreliance, and underreliance on an AI chatbot.

\section{Results}

\subsection{RQ1: Factors Associated with Appropriate Reliance on AI}

Our analysis revealed several pre-survey factors significantly associated with students' appropriate reliance on AI systems. Programming self-efficacy showed a positive correlation with appropriate reliance ($r = 0.333$, $p = .018$), indicating that students with higher confidence in their programming abilities were more likely to rely on AI appropriately. Similarly, programming literacy demonstrated a significant positive correlation with appropriate reliance ($r = 0.370$, $p = .008$), suggesting that higher domain knowledge helps students discern when to trust AI recommendations. Need for cognition was positively correlated with appropriate reliance ($r = 0.371$, $p = .008$) as well.

Post-task measures had significant relationships with appropriate reliance. Students' decision-making processes were positively correlated with appropriate reliance ($r = 0.312$, $p = .027$), suggesting that specific approaches to problem-solving support more effective AI use. Trust-related outcomes showed negative correlations with appropriate reliance. Final trust in the chatbot was negatively correlated with appropriate reliance ($r = -0.303$, $p = .032$), as was satisfaction  ($r = -0.292$, $p = .040$). This suggests that students who developed a more skeptical stance toward the AI system were more likely to use it appropriately.

\subsection{RQ2: Factors Associated with Overreliance on AI} 
None of the pre-survey measures (programming self-efficacy, general trust in AI, concerns for trust, need for cognition, programming literacy, and AI literacy) showed significant correlations with overreliance. This contrasts with our findings for appropriate reliance, where several pre-survey factors showed significant relationships. One possible explanation is that overreliance is more dependent on the interaction experience itself rather than from pre-existing traits or knowledge. Participants may approach AI recommendations with initial caution, but their decision to overrely develops during the interaction process. 

However, post-task measures showed significant relationships with overreliance. Final trust in the chatbot demonstrated a strong positive correlation with overreliance ($r = 0.441$, $p = .001$), as did satisfaction ($r = 0.348$, $p = .013$). Students' self-assessment was somewhat accurate, as post-task overreliance self-assessment positively correlated with students' measured overreliance behavior ($r = 0.355$, $p = .011$). Additionally, students' decision-making processes showed a negative correlation with overreliance ($r = -0.265$, $p = .063$), though this relationship fell just short of statistical significance at the $p < .05$ threshold.

\subsection{RQ3: Factors Associated with Underreliance on AI}

Our analysis of factors related to underreliance on AI revealed several significant relationships. Among pre-survey measures, programming self-efficacy showed a negative correlation with underreliance ($r = -0.308$, $p = .030$), indicating that students with higher confidence in their programming abilities were less likely to reject helpful AI recommendations. Similarly, need for cognition was negatively correlated with underreliance ($r = -0.312$, $p = .027$), suggesting that students who enjoy thinking deeply about problems were more likely to recognize helpful AI recommendations. The strongest negative correlation with underreliance was observed for programming literacy ($r = -0.439$, $p = .001$), indicating that students with stronger programming knowledge were significantly less likely to dismiss helpful AI recommendations. Other pre-survey measures, including general trust in AI, concerns for trust, and AI literacy, did not show significant correlations with underreliance. Similarly, none of the post-task measures (final overreliance assessment, final trust in chatbot, satisfaction with AI, and decision-making process) were significantly correlated with underreliance.

\section{Discussion}

This study sought to understand what individual characteristics predict different patterns of AI reliance among undergraduate students. Our findings reveal several factors associated with appropriate reliance, overreliance, and underreliance behaviors. We discuss three insights relating to the findings and their implications.

First, our findings identify a clear profile of students who use AI effectively: those with higher programming self-efficacy (r = 0.333), programming literacy (r = 0.370), and need for cognition (r = 0.371). These students demonstrated the ability to correctly accept helpful AI recommendations while rejecting misleading ones, indicating calibrated trust and thoughtful evaluation. This finding aligns with Gaube et al. \cite{gaube2021ai} and Li and Little's \cite{li2023appropriate} observations, where professionals with domain expertise are less vulnerable to uncritically accepting flawed AI recommendations. In our educational context, programming literacy functions as this protective factor. Students with higher programming literacy appear to leverage their understanding of coding concepts to evaluate the AI recommendations in the study. The role of need for cognition complements this domain expertise by ensuring students actively scrutinize rather than passively accept AI suggestions, as was found in \cite{de2025cognitive,buccinca2021trust,vasconcelos2023explanations}. Similarly, self-efficacy provides students the confidence necessary to trust one's own judgment when AI recommendations seem questionable. These factors distinguish students who appropriately rely on AI, suggesting that educational interventions should focus on developing domain expertise, critical thinking skills, and the confidence to question AI outputs.

Second, our analysis reveals a concerning relationship between trust, satisfaction, and overreliance during students' interactions with AI. While no pre-task measures predicted overreliance, post-task trust (r = 0.441) and satisfaction (r = 0.348) in the AI chatbot showed strong positive correlations with overreliance. These correlations indicate that overreliant students trusted the system, felt satisfied with its assistance, yet unknowingly accepted errors that undermined their performance. This finding provides evidence for a potential "false-confidence loop" in human-AI interactions. Unlike human advisors, whose errors might trigger skepticism through behavioral cues or inconsistencies, AI systems can deliver incorrect information with confidence.

Students who develop trust through initial positive experiences appear to generalize this trust inappropriately, leading them to accept subsequent recommendations with less scrutiny. Each uncritically accepted suggestion reinforces their satisfaction with the AI, which in turn strengthens their trust, creating a self-perpetuating cycle. This extends automation bias research by demonstrating how the very features that make AI appealing (consistency, availability, and authoritative presentation) not only prevent users from recognizing limitations but actively reinforce overreliant behavior patterns, creating a feedback loop where each accepted recommendation deepens trust in the system.

Third, our finding that students can partially recognize their own overreliance (r = 0.355, p = .011) represents an important opportunity for intervention design. This significant correlation suggests students possess some metacognitive awareness of when they're accepting AI recommendations too readily. While this self-awareness isn't perfect, it provides a foundation that educational interventions can build upon. For instance, interventions could prompt students to predict their own reliance patterns during AI interactions and reflect on how they engage with the outputs, helping calibrate self-perception with reality. This aligns with cognitive forcing functions research \cite{buccinca2021trust}, which found that prompting users to think before accepting AI recommendations reduces overreliance. Our finding that students possess partial self-awareness suggests they would be receptive to such interventions that make their reliance patterns explicit. By leveraging students' existing metacognitive awareness through reflection prompts, acceptance rate feedback, or required justifications for following AI suggestions, we can strengthen their ability to recognize and correct overreliant behaviors.

\section{Limitations and Future Work}

This pilot study has several limitations that suggest directions for future research. Our sample of computer science students may not generalize to other disciplines where domain expertise operates differently. Students in fields like humanities or social sciences might show different reliance patterns when using AI for writing or analysis tasks. Additionally, our experimental design using pre-programmed AI responses, while ensuring consistency, may not capture the variability of real-world AI systems that students encounter.

Our measurement of AI literacy may also require refinement. The lack of correlation between AI literacy and reliance patterns could indicate that students' self-reported AI knowledge reflects personal experience and technical understanding rather than their ability to critically assess AI recommendations. Students may feel knowledgeable about AI while lacking skills to identify unreliable outputs.
Future work should develop AI literacy assessments and educational curriculum focused on practical competencies (e.g. recognizing hallucinations, understanding when AI presents uncertain information as fact, and identifying when AI assistance may be harmful to learning).

Furthermore, examining reliance patterns within a single session prevented us from understanding how these behaviors develop and solidify over time. The "false-confidence loop" we identified may strengthen or weaken with extended use. Future research should employ longitudinal designs to track how reliance patterns change over time, test intervention strategies like embedded reflection prompts, and examine whether the related factors we identified remain stable across extended AI use. 

\section{Conclusion}

This study investigated factors contributing to different patterns of AI reliance among undergraduate students. Through a controlled experiment where students solved programming problems with an AI assistant providing both accurate and misleading recommendations, we identified distinct predictors for appropriate reliance, overreliance, and underreliance behaviors. Students who demonstrated appropriate reliance possessed higher programming self-efficacy, domain expertise, and need for cognition. In contrast, overreliance showed no correlation with pre-task measures but was strongly associated with high post-task trust and satisfaction. Underreliance was negatively associated with domain expertise and need for cognition. These findings highlight the need for educational approaches that build domain expertise, foster critical thinking skills, and incorporate reflection mechanisms to help students calibrate their trust in AI systems. The partial self-awareness students demonstrated about their reliance patterns in the post-survey provides an opportunity for interventions that make inappropriately reliant behaviors known and correctable. As AI becomes increasingly integrated in education, understanding and addressing patterns of under- and overreliance will be necessary for helping students develop the judgment to leverage these tools effectively without compromising their ability to learn and think independently.

\bibliographystyle{splncs04}
\bibliography{ref}
\end{document}